\documentclass[aps,prb,twocolumn,reprint,showpacs]{revtex4-1}
\usepackage{amsmath}
\usepackage{amssymb}
\usepackage{graphicx}
\usepackage{hyperref}
\begin{document}
\title{Charge transfer and hybrid ferroelectricity in (YFeO$_{3}$)$_{n}$/(YTiO$_{3}$)$_{n}$ magnetic superlattices}
\author{Huimin Zhang}
\author{Yakui Weng}
\author{Xiaoyan Yao}
\author{Shuai Dong}
\email{sdong@seu.edu.cn}
\affiliation{Department of Physics, Southeast University, Nanjing 211189, China}
\date{\today}

\begin{abstract}
Interfaces in oxide heterostructures always provide a fertile ground for emergent properties. Charge transfer from a high energy band to a low energy opponent is naturally expected, as occurring in semiconductor $p$-$n$ junctions. In this study, several exceptional physical phenomena have been predicted in (YFeO$_3$)$_n$/(YTiO$_3$)$_n$ superlattices. First, the charge transfer between these Mott insulators is in opposite to the intuitive band alignment scenario. Second, hybrid ferroelectricity with a moderate polarization is generated in the $n=2$ magnetic superlattice. Furthermore, the ferroelectric-type distortion can persist even if the ($AB$O$_3$)$_2$/($AB$'O$_3$)$_2$ system turns to be metallic, rending possible metallic ferroelectricity.
\end{abstract}
\maketitle

\section{Introduction}
Oxide heterostructures provide a unique and fertile ground to study emergent physics of correlated electrons and a promising route to design new devices using quantum effects. At the interfaces between oxides, collective behaviors of electrons can be contrastively different from their original roles in parent materials \cite{Zubko:Arcmp,Hwang:Nm}. For example, previous studies have revealed plethoric phenomena relevant to electronic reconstructions, e.g. metal-insulator transition and enhanced N\'{e}el temperatures in LaMnO$_{3}$/SrMnO$_{3}$ superlattices \cite{Dong:Prb08.3,May:Nm,Nanda:Prb,Adamo:Prb}, two-dimensional electronic gas and superconductivity in LaAlO$_{3}$/SrTiO$_{3}$ heterostructures \cite{Ohtomo:Nat04,Nakagawa:Nm}, orientation-dependent magnetism in LaNiO$_{3}$/LaMnO$_{3}$ \cite{Gibert:Nm,Dong:Prb13} and LaFeO$_{3}$/LaCrO$_{3}$ \cite{Ueda:Sci,Zhu:Jap} superlattices.

Charge transfer is one important driving force for electronic reconstructions, which tunes the local electron density as well as the interfacial electrostatic field. In the light of the band alignment scenario, the heights of Fermi levels are the decisive factor for charge transfer. Naturally, to reach a uniform chemical potential across the interface, electrons will leak from the high energy band to the low energy opponent, which is well known in semiconductor $p$-$n$ junctions.

Different from the nearly-free electrons, the bands of correlated electrons are not strictly rigid but somewhat ``soft" \cite{Kotliar:Pto,Kawasugi:Prl,Dagotto:Sci}, which may lead to emergent phenomena beyond the simple band theory. In this work, the (YFeO$_{3}$)$_{n}$/(YTiO$_{3}$)$_{n}$ superlattices have been studied using the density functional theory (DFT). Several unexpected physical phenomena, e.g. the charge transfer against the band alignment scenario and hybrid multiferroicity, have been predicted. In addition, due to the origin of improper component of polarization, the ferroelectric-type distortion in ($AB$O$_3$)$_2$/($AB$'O$_3$)$_2$ superlattices can persist even if the system turns to be metallic.

\section{Model \& method}
Both YFeO$_{3}$ and YTiO$_{3}$ are Mott insulators \cite{Mochizuki:Njp,Pasternak:Mrsp}. By neglecting the weak ferromagnetism due to tiny spin canting, the magnetic ground state of YFeO$_{3}$ is antiferromagnetic (AFM) \cite{Pasternak:Mrsp,Alaria:Cs}. In contrast, YTiO$_{3}$ is ferromagnetic (FM) \cite{Mochizuki:Njp}. These two materials share the common A-site cation Y$^{3+}$, as well as the identical space group of crystal structure (orthorhombic $Pbnm$, see Fig.~\ref{bulk}(a)). Their lattice constants ($a$, $b$, $c$) in unit {\AA} are: ($5.283$, $5.592$, $7.603$) for YFeO$_{3}$ \cite{Treves:Jap} and ($5.338$, $5.690$, $7.613$) for YTiO$_{3}$ \cite{Hester:Acb}. The proximate structures allow a high possibility for epitaxial growth of multilayers. In the following, the (YFeO$_{3}$)$_{n}$/(YTiO$_{3}$)$_{n}$ superlattices are assumed to be grown on the mostly-used SrTiO$_{3}$ ($001$) substrate. To match the substrate, the in-plane lattice constants of YFeO$_{3}$ and YTiO$_{3}$ are fixed as $3.905\times\sqrt{2}=5.5225$ {\AA}.

All the following calculations are performed using the Vienna \textit{ab initio} Simulation Package (VASP) \cite{Kresse:Prb,Kresse:Prb96} based on the generalized gradient approximation (GGA). The Hubbard $U_{\rm eff}$ ($=U-J$) is imposed on Fe's and Ti's $d$ orbitals using the Dudarev implementation \cite{Dudarev:Prb}. In the GGA+$U$ calculation, the plane-wave cutoff is $550$ eV. A $7\times7\times5$ Monkhorst-Pack $k$-point mesh centered at $\Gamma$ point is adopted for (YFeO$_{3}$)$_{1}$/(YTiO$_{3}$)$_{1}$ and the parent materials, while it is $6\times6\times2$ for (YFeO$_{3}$)$_{2}$/(YTiO$_{3}$)$_{2}$. Both the lattice constant along the $c$-axis and inner atomic positions are fully relaxed till the Hellman-Feynman forces are all below $0.01$ eV/{\AA}.

 For comparison, the hybrid functional calculations based on the Heyd-Scuseria-Ernzerhof (HSE) exchange \cite{Heyd:Jcp,Heyd:Jcp04,Heyd:Jcp06} are also performed. Due to its extreme demand of CPU-time, the plane-wave cutoff is reduced to $400$ eV. And the $k$-point mesh is reduced to $3\times3\times2$ for (YFeO$_{3}$)$_{1}$/(YTiO$_{3}$)$_{1}$ and the parent materials, while it is $3\times3\times1$ for the $n=2$ superlattice.

\section{Results \& discussion}
\subsection{Band alignment}
Before the study on superlattices, the parent materials are checked. According to the previous literature \cite{Alaria:Cs,Huang:Jap}, proper $U_{\rm eff}$ values $U_{\rm Fe}=4$ eV and $U_{\rm Ti}=3.2$ eV are adopted in the following, if not noted explicitly. In our calculation, the magnetic ground states of YFeO$_{3}$ and YTiO$_{3}$ are G-type AFM (G-AFM) and FM, respectively. Both Fe$^{3+}$ and Ti$^{3+}$ are in the high-spin states. The relaxed lattice constants also agree with the experimental values quite well \cite{Treves:Jap,Hester:Acb}. All these results guarantee the reliability of following calculations on superlattices.

\begin{figure*}
\centering
\includegraphics[width=0.25\textwidth]{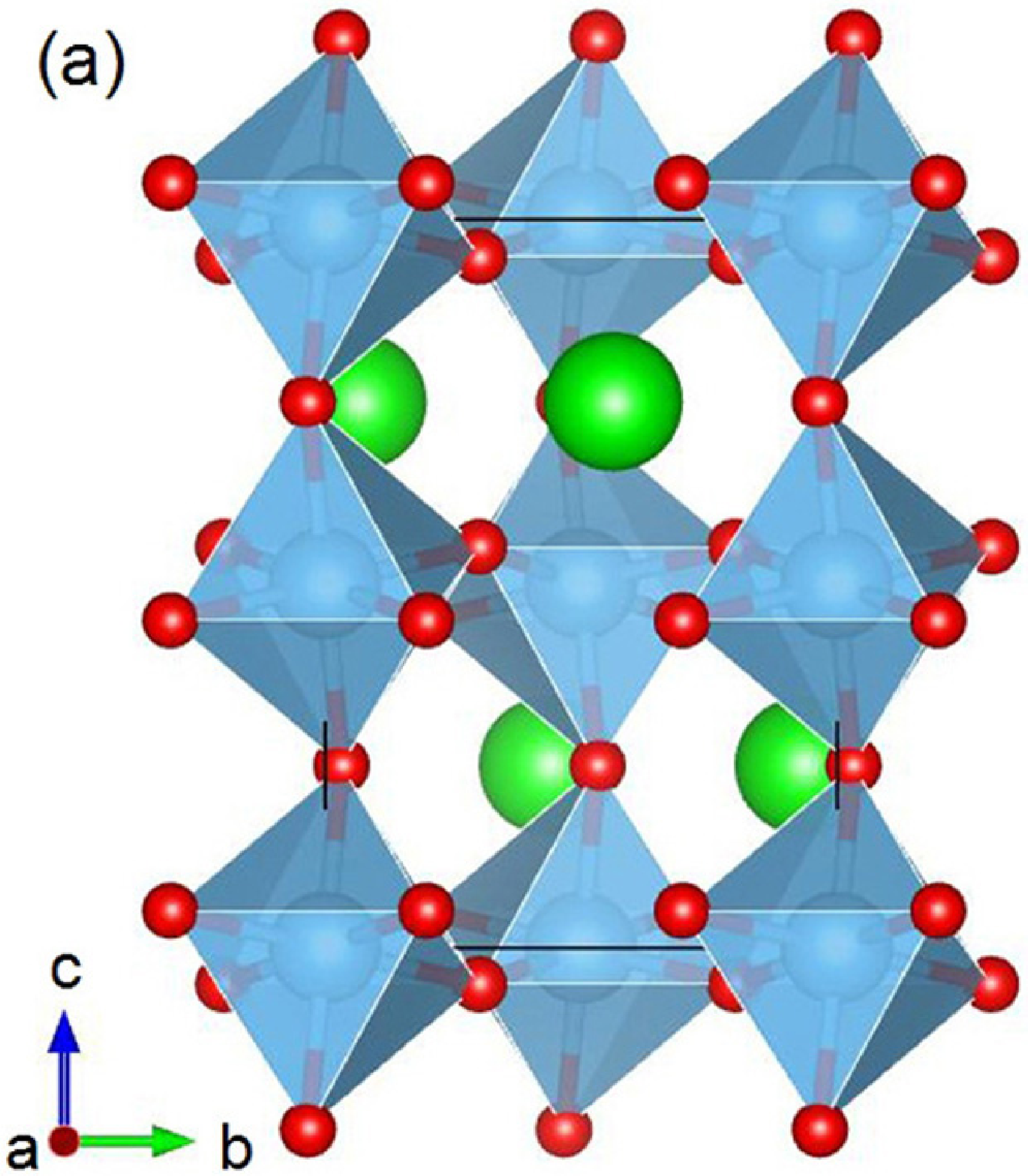}\includegraphics[width=0.31\textwidth]{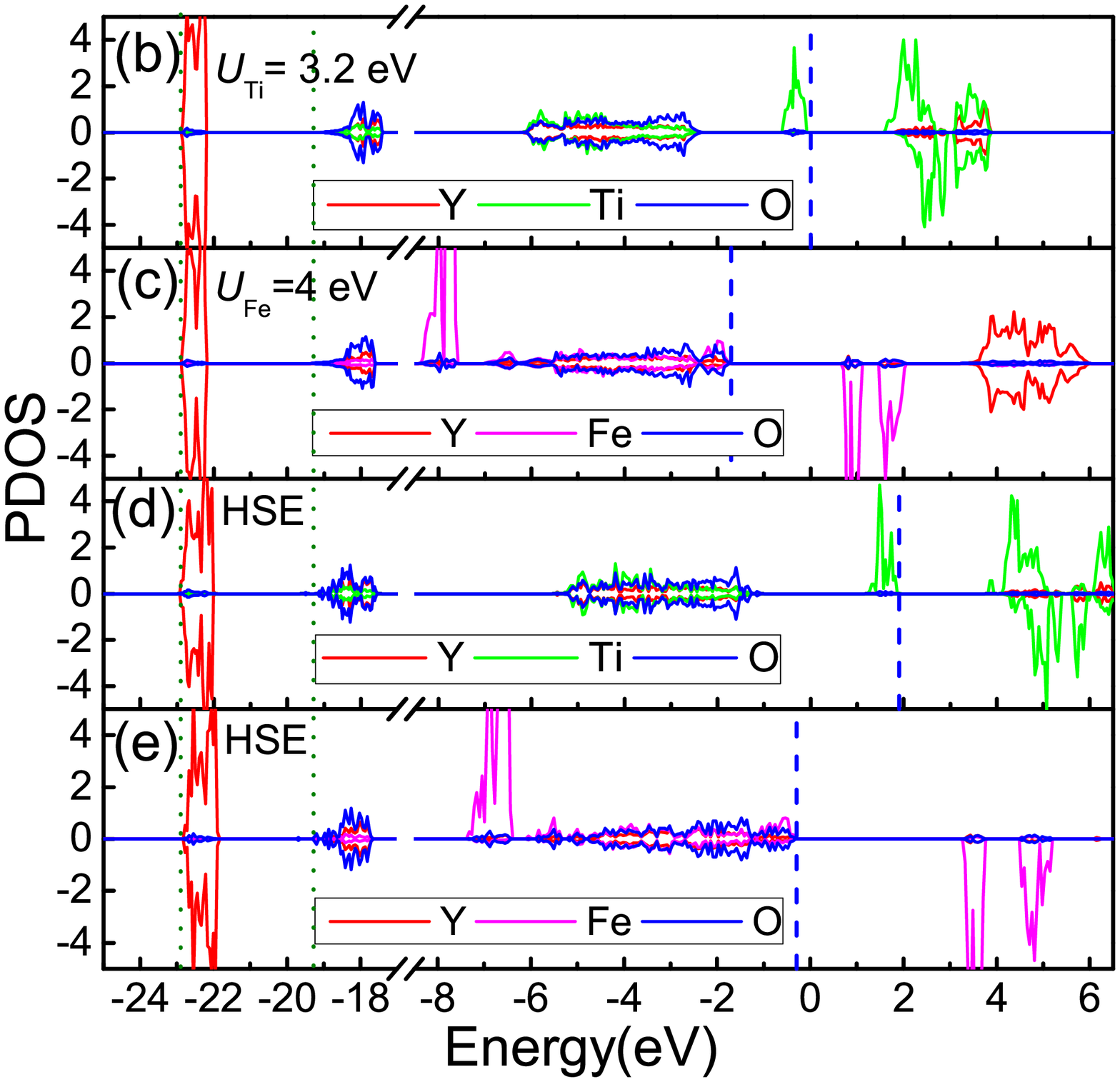}\includegraphics[width=0.42\textwidth]{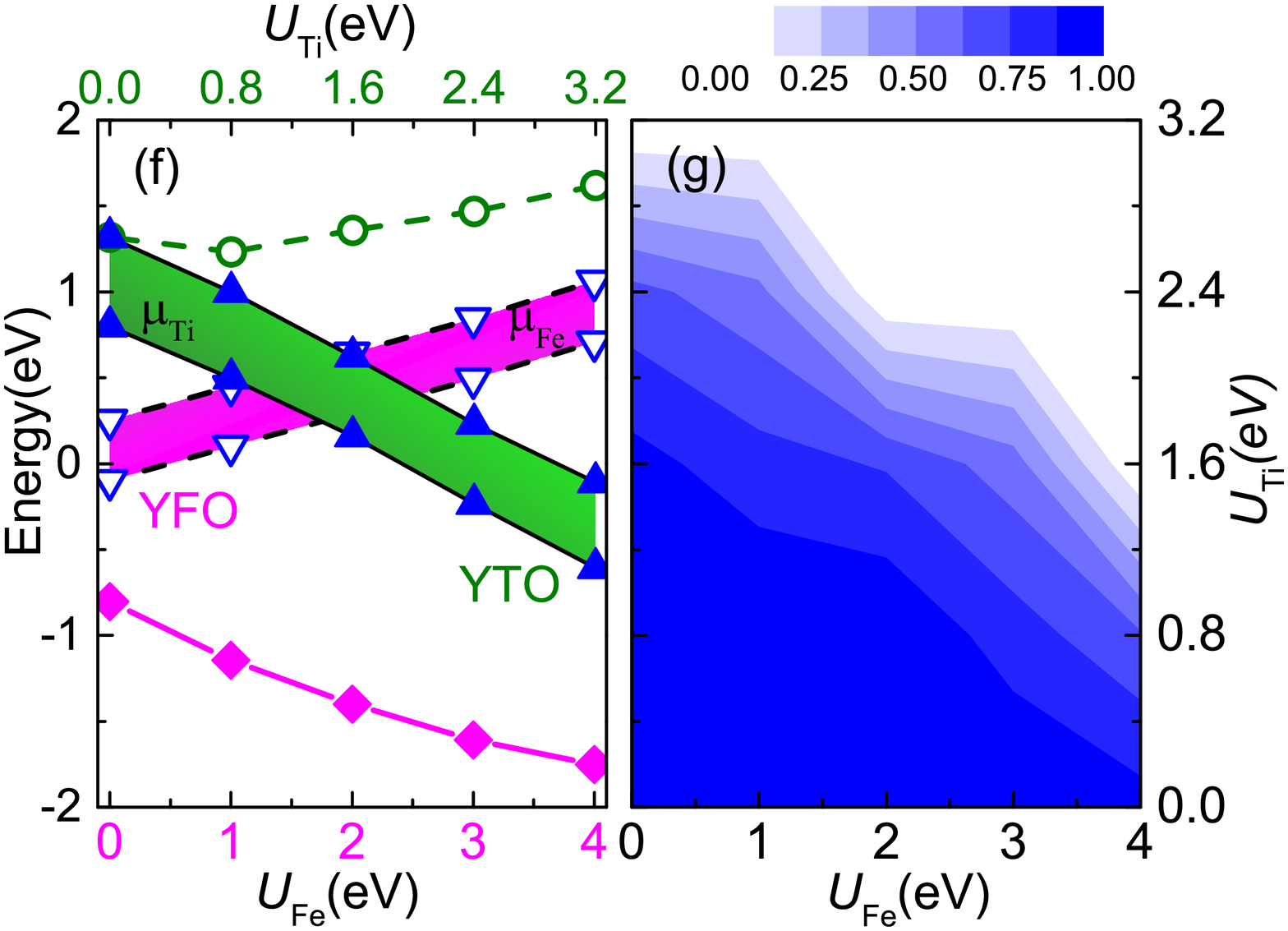}
\caption{(Color online) Properties of YTiO$_{3}$ and YFeO$_{3}$. (a) Sketch of the common structure. (b-e) Projected density of states (PDOS) for YTiO$_{3}$ [(b) and (d)] and YFeO$_{3}$ [(c) and (e)]. The Fermi level in (b) is set as zero and the deep energy bands of Y's $4p$ (around $-23$ to $-22$ eV) and O's $2s$ orbitals (around $-19$ to $-17.5$ eV) are aligned for all four PDOS's, as denoted by (green) dot lines. The Fermi level for each case is also marked by a (blue) broken line. (b-c) Obtained using GGA+$U$. (d-e) Obtained using HSE. (f) After the treatment of band alignment, the energy positions of near-Fermi-level bands of YTiO$_3$ (green upper-triangles, $\mu_{\rm Ti}$) and YFeO$_3$ (pink lower-triangles, $\mu_{\rm Fe}$) as functions of $U_{\rm Ti}$ (upper horizontal axis) or $U_{\rm Fe}$ (lower horizontal axis). In addition, the band edges of 1) topmost valence band of YFeO$_3$ (pink rhombs) and 2) bottommost conducting band of YTiO$_3$ (green circles) are also presented. Occupied/empty bands are marked by solid/open symbols, respectively. (g) Contour of charge transfer as a function of $U_{\rm Fe}$ and $U_{\rm Ti}$ according to the band alignment. The numerical label is on the top of box.}
\label{bulk}
\end{figure*}

Then the substrate strain is imposed. The G-AFM order persists in strained YFeO$_{3}$ on SrTiO$_{3}$ substrate. However, for YTiO$_{3}$, the strain can drive a magnetic transition to A-type antiferromagnet \cite{Zhou:Jap}. The atomic projected density of states (PDOS) of strained YFeO$_{3}$ and YTiO$_{3}$ are displayed in Fig.~\ref{bulk}(b) and \ref{bulk}(c), respectively. Both materials retain insulating with energy gaps of $\sim2.4$ eV for strained YFeO$_{3}$ and $\sim1.5$ eV for strained YTiO$_{3}$.

According to Fig.~\ref{bulk}(b-c), the topmost valence band of YTiO$_{3}$ is from Ti's one $t_{\rm 2g}$ orbital (whose position is denoted as $\mu_{\rm Ti}$), and the bottommost conducting band of YFeO$_{3}$ is formed by the spin-down $t_{\rm 2g}$ orbitals of Fe (whose position is denoted as $\mu_{\rm Fe}$). Both these two bands are very narrow, implying localized states. By aligning the deep energy bands of Y's $4p$ and O's $2s$ orbitals which should be identical in these two materials, a band alignment can be obtained straightforwardly. Ti's occupied $t_{\rm 2g}$ band just locates within the forbidden gap of YFeO$_{3}$, lower than the unoccupied conducting band of YFeO$_{3}$. Mathematically, it can be expressed as $\mu_{\rm Ti}<\mu_{\rm Fe}$. Thus an intuitive conclusion is that the charge transfer should be forbidden between these two materials, keeping the original Fe$^{3+}$-Ti$^{3+}$ configuration across the interface. This band alignment is further confirmed in the HSE calculations (Fig.~\ref{bulk}(d-e)).

Even if the above $U_{\rm Fe}=4$ eV and $U_{\rm Ti}=3.2$ eV are believed to be the most appropriate, it is interesting to go beyond the real materials by scanning the Hubbard $U_{\rm eff}$ in a wider parameter space. By varying $U_{\rm eff}$'s, the aligned band positions are summarized in Fig.~\ref{bulk}(f). In general, $\mu_{\rm Ti}$ decreases with increasing $U_{\rm eff}$, while $\mu_{\rm Fe}$ increases with increasing $U_{\rm eff}$. This tendency can be understood based on the Hubbard model, since the occupied Ti's $t_{\rm 2g}$ band is the lower-Hubbard band while the unoccupied Fe's band is the upper-Hubbard one. Then, the intensity of charge transfer can be determined by comparing $\mu_{\rm Ti}$ and $\mu_{\rm Fe}$, as summarized in Fig.~\ref{bulk}(g). When both $U_{\rm Ti}$ and $U_{\rm Fe}$ are large, there is no charge transfer, as illustrated in Fig.~\ref{bulk}(b-c). In contrast, while in the small $U_{\rm Ti}$ and $U_{\rm Fe}$ limit, a partial or complete charge transfer should occur.

\subsection{Charge transfer}
\begin{figure}
\centering
\includegraphics[width=0.48\textwidth]{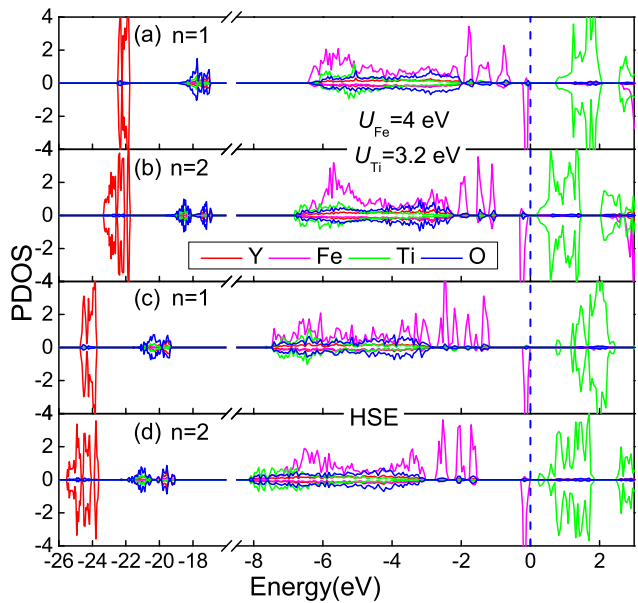}
\caption{(Color online) PDOS of superlattices. (a-b) GGA+$U$. (c-d) HSE. The Fermi level is positioned at zero (blue broken line). For $n=1$, the deep Y's $4p$ and O's $2s$ bands are similar to their bulk's correspondences, while for $n=2$, clear splittings are observed due to the electrostatic potential created by the charge transfer.}
\label{sl}
\end{figure}

Above analyses on charge transfer were based on the band alignment scenario. To verify this scenario, (YFeO$_{3}$)$_{n}$/(YTiO$_{3}$)$_{n}$ superlattices ($n=1$ and $2$) are studied. After relaxing the crystal structures, the magnetic ground states of superlattices are determined. In both superlattices, the magnetic moment of Ti is (almost) quenched, while the YFeO$_{3}$ layers retain the G-AFM configuration with a suppressed moment $\sim3.6$ $\mu_{\rm B}$/Fe. This magnetic reconstruction is due to the charge transfer, since Ti$^{4+}$ is non-magnetic and the moment of high-spin Fe$^{2+}$ is $4$ $\mu_{\rm B}$/Fe. This scenario is further confirmed by the PDOS's (Fig.~\ref{sl}(a-b)). One of Fe's upper Hubbard bands is occupied by one electron, while Ti's $3d$ bands are fully empty, implying a complete charge transfer. Such a complete charge transfer is further confirmed using the HSE calculation (Fig.~\ref{sl}(c) and (d)).

For physical comparison, the GGA+$U$ calculations with lower $U_{\rm eff}$'s have also been done. The charge transfer always occurs in superlattices' calculation despite the value of $U_{\rm eff}$'s. In other words, the prediction of charge transfer is not $U_{\rm eff}$-sensitive. In addition, our GGA+$U$ and HSE calculations give consistent results regarding the charge transfer, which are also in agreement with the previous LDA/LDA+$U$ calculations and X-ray photoemission experiment on similar system LaTiO$_3$/LaFeO$_3$ superlattices \cite{Kleibeuker:Prl}. Thus the prediction is not an artefact of the level of treatment of electronic correlations.

Then how to understand such an unexpected charge transfer, which violates the band alignment scenario? Previously, similar charge transfer in LaTiO$_3$/LaFeO$_3$ superlattices was attributed to the transition of Fe$^{2+}$ from the high-spin state to low-spin state \cite{Kleibeuker:Prl}. However, in our case, Fe$^{2+}$ remains in the high-spin state, giving $\sim4$ $\mu_B$ per Fe. In this sense, such a theoretical argument based on the low-spin state can not interpret the ``anomalous" charge transfer predicted here.

One may suspect that such an unexpected charge transfer is due to the reduced dimension of YFeO$_{3}$ and YTiO$_{3}$ layers in superlattices. The band alignment shown in Fig.~\ref{bulk}(b-c) is obtained according to the three-dimensional bulks, which may be reshaped into ultra-thin layers. Taking a tight-binding model for illustration, the reduced dimension can only tune the bandwidths but not the (central) positions of bands. According to Fig.~\ref{bulk}(f), the shrinking of bandwidths can not reverse the band alignment when $U_{\rm Ti}=3.2$ eV and $U_{\rm Fe}=4$ eV. Thus, pure dimension reduction is not enough.

Dimension reduction can also tune the Mottness by tuning the ratio of bandwidth and Hubbard $U$. Then the splitting between the upper and lower Hubbard bands can be shifted, which may alter the band alignment. However, this possibility can also be simply ruled out. The dimension reduction can only suppress the kinetic energy and thus prefers the Mottness, equivalent to increase Hubbard $U$. According to Fig.~\ref{bulk}(f), the charge transfer is suppressed to zero with enhanced $U$. In short, the dimension reduction can not explain the unexpected charge transfer.

All above evidences suggest that the reconstruction of Hubbard bands is the intrinsic origin for such an ``anomalous" charge transfer. As illustrated in Fig.~\ref{sl}, for both Ti and Fe, the energy splittings between upper and lower Hubbard bands become much smaller than those in parent materials. The Fe$^{3+}$, with half-filling $3d$ orbitals, is the most prominent candidate for the Hubbard repulsion between the spin-up and spin-down bands. However, the Hubbard repulsion will be suppressed when the electron density deviates from the half-filling, as in Fe$^{2+}$. Similar mechanism works for Ti ions. In other words, the bands of YFeO$_{3}$ and YTiO$_{3}$ are not rigid neither but rather fragile against (virtual) hopping of electrons. Namely, although $\mu_{\rm Fe}>\mu_{\rm Ti}$, both $\partial\mu_{\rm Fe}/\partial n$ and $\partial\mu_{\rm Ti}/\partial n$ are negative, where $\partial n$ denotes charge transfer from Ti to Fe.

Similar mechanism should also work in LaFeO$_3$/LaTiO$_3$, which induces the charge transfer and low-spin state of Fe$^{2+}$ \cite{Kleibeuker:Prl}. The low-spin state of Fe$^{2+}$ does not occur in our calculation due to the structural difference between LaFeO$_3$ and YFeO$_3$. Following Ref.~\onlinecite{Kleibeuker:Prl}, we can reproduce the transition from the high-spin state to low-spin state in LaFeO$_3$ by adding one more electron to Fe. However, the same treatment shows that the high-spin state of YFeO$_3$ is more stable. The highly distorted structure of YFeO$_3$ reduces the bandwidth of $3d$ orbitals, which prefers the Mottness and Hubbard splitting between spin-up and spin-down. In fact, the high-pressure experiment also found the high-spin state of orthoferrites is more stable when the A-site ion is small \cite{Pasternak:Mrsp}.

Finally, it should be noted that although the Fe$^{2+}$-Ti$^{4+}$ configuration also occurs in other systems, e.g. FeTiO$_3$ and Ti doped Fe$_2$O$_3$ \cite{Harrison:Nat,Velev:Prb,Pentcheva:Prb}, the involved mechanism may be not simply identical. Although the electronegativity of different ions usually plays a crucial role \cite{Chen:Prl}, other factors, especially the coordination environment and correlation, will also tune the charge transfer especially when the alignment between involved bands is subtle, as demonstrated in the present work.

\subsection{Hybrid ferroelectricity}
It was predicted that improper ferroelectricity could emerge in perovskite superlattices like (PbTiO$_{3}$)$_{1}$/(SrTiO$_{3}$)$_{1}$ \cite{Bousquet:Nat}, (LaFeO$_{3}$)$_{n}$/(YFeO$_{3}$)$_{n}$ \cite{Alaria:Cs}, as well as the layer perovskites $A_3B_2$O$_7$ \cite{Benedek:Prl,Mulder:Afm}, due to the modulation of non-polar antiferrodistortive modes. For these ($AB$O$_{3}$)$_{n}$/($A'B$O$_{3}$)$_{n}$, the ferroelectric polarization appears to be nonzero when $n$ is an odd number, while it is fully compensated between layers for even $n$'s \cite{Rondinelli:AM12}.

Our superlattices, in the type of ($AB$O$_{3}$)$_{n}$/($AB'$O$_{3}$)$_{n}$, is somewhat different. First, the crystal structures are analyzed. The space group of $n=1$ superlattice is $P2_{1}/c$ , but that of $n=2$ is $Pmc2_{1}$. The point group of $P2_{1}/c$ is $2/m$, and $mm2$ for $Pmc2_{1}$. $2/m$ belongs to the non-polar point group while $mm2$ is a polar point group with its polar axis along the $b$-axis. Therefore, the $n=1$ superlattice should be non-ferroelectric while a finite polarization is expectable in the $n=2$ superlattice. Such a prediction based on symmetry analysis should be physical robust and qualitatively independent on details of numerical calculations (e.g. $U_{\rm eff}$).

The standard Berry phase method implemented in VASP is employed to evaluate the ferroelectric polarization. As expected, the calculated polarization is zero for the $n=1$ superlattice, but in the case of $n=2$, a net polarization up to $\sim1.01$ $\mu$C/cm$^{2}$ is obtained along the $b$-axis. These results can be verified using piezoelectric force microscopy or optical second harmonic generation, as done in the LaFeO$_3$/YFeO$_3$ superlattices \cite{Alaria:Cs}. In addition, the intuitive point charge model can also be employed to estimate the approximate value of polarization, which gives $\sim2.1$ $\mu$C/cm$^{2}$, in agreement with the corresponding Berry phase values qualitatively. The quantitative difference is reasonable, since the Born effective charge of ions can be different from their nominal valences.

The ferroelectricity, together with the magnetic ordering of Fe, makes the $n=2$ YFeO$_3$/YTiO$_3$ superlattice multiferroic. It is noteworthy that the previously studied $n=2$ LaFeO$_3$/LaTiO$_3$ superlattice is non-magnetic due to the low-spin state of Fe \cite{Kleibeuker:Prl}. In this sense, the YFeO$_3$/YTiO$_3$ superlattices can provide more physical functions.

\begin{figure}
\centering
\includegraphics[width=0.45\textwidth]{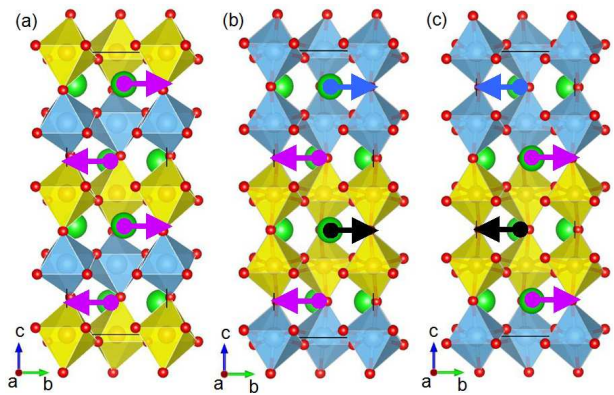}
\caption{(Color online) Sketch of ferroelectric distortions. The arrows denote the displacements of Y$^{3+}$. (a) $n=1$. The displacements are compensated between layers. The (b) positive and (c) negative ferroelectric distortion for $n=2$.}
\label{distortion}
\end{figure}

The origin of polarization can be visualized in Fig.~\ref{distortion}. Due to the antiferrodistortive mode, both Y$^{3+}$ and O$^{2-}$ will move away from their corresponding high-symmetric positions in the high-temperature cubic structure. In the $n=1$ superlattice, all displacements have their asymmetric opponents nearby (Fig.~\ref{distortion}~(a)), thus can not generate a net polarization, as in the parent materials. In contrast, in the $n=2$ superlattice, the sequence of B-site cations forms ...-Ti-Ti-Fe-Fe-... along the $c$-axis. This sequence is analogous to the ...-$\uparrow$-$\uparrow$-$\downarrow$-$\downarrow$-... spin structure in some type-II multiferroics (e.g. Ca$_{3}$CoMnO$_{6}$ \cite{Choi:Prl}, o-HoMnO$_3$ \cite{Sergienko:Prl,Picozzi:Prl}, and BaFe$_{2}$Se$_{3}$ \cite{Dong:Prl14}), which owns parity. This parity, together with the antiferrodistortive mode, breaks the space inversion symmetry \cite{Dong:Prl14}. The displacements of Y$^{3+}$ (and O$^{2-}$) sandwiched between the Ti-Ti (or Fe-Fe) bilayers and Ti-Fe bilayers are no longer asymmetric, giving rise to a net in-plane polarization. Such distortions can be reversed by reversing the antiferrodistortive pattern, as illustrated in Fig.~\ref{distortion}(b) and \ref{distortion}(c). Then the ferroelectric polarization is switched from positive to negative, as confirmed in the Berry phase calculation.

Considering the similarity and difference, our result on ($AB$O$_{3}$)$_{n}$/($AB$'O$_{3}$)$_{n}$ is complementary to the previous studied ($AB$O$_{3}$)$_{n}$/($A$'$B$O$_{3}$)$_{n}$, completing the theory of improper ferroelectricity in perovskite superlattices driven by the modulation of non-polar antiferrodistortive mode \cite{Rondinelli:AM12}. In fact, the parity-related origin of ferroelectricity in ($AB$O$_{3}$)$_{n}$/($AB$'O$_{3}$)$_{n}$ superlattice is even more general, which can be independent of the details of antiferrodistortive mode.

In the point charge model, a semi-quantitative partition can be applied to the total ferroelectric polarization to extract various contributions. The ferroelectric contributions of Ti-O-octahedra, Fe-O-octahedra, and Y-O-icosahedra (Fig.~\ref{fe}(a)) can be estimated as \footnote{To analyze the origin of ferroelectricity, the net dipole moment is partitioned into three components from Fe-O-octahedra, Ti-O-octahedra, and Y-O-icosahedra using the point charge model. First, the centroid of each oxygen octahedron $\vec{r}_{\rm O}$ is calculated. Then the dipole moments $\vec{P}_{\rm Fe-O}$ and $\vec{P}_{\rm Ti-O}$ are calculated using $eV_X(\vec{r}_X-\vec{r}_{\rm O})$, where $V_X$ ($X$=Fe and Ti respectively) denotes the valence of cation and $e$ is the elementary charge. In this calculation, the oxygen octahedra only contribute the equivalent negative charge $-eV_X$, while the rest negative charge forms dipole moments together with Y$^{3+}$. This method can estimate the individual contribution to total polarization semi-quantitatively.}: $-144\%$, $+82\%$, and $+162\%$ to the net polarization, as shown in Fig.~\ref{fe}(b). It is clear that the Y-O-icosahedra contribute the most polarization, which is partially compensated by the Ti-O-octahedra one. The large negative one from Ti-O-octahedra can be considered as the induced one polarized by the inner-built field from Y-O-icosahedra's ferroelectricity since Ti$^{4+}$ is a proper ferroelectric active ion with a considerable large dielectric coefficient. In this sense, the ferroelectric polarization in our $n=2$ superlattice is not a pure improper one as in the previous studied cases \cite{Alaria:Cs,Benedek:Prl,Mulder:Afm}, but a hybrid one with both improper and proper contributions.

\subsection{Possible metallic ferroelectricity}
Very recently, a few studies found some peculiar materials, in which the ferroelectric distortion persists despite its metallicity \cite{Shi:Nm,Puggioni:Nc,Xiang:Prb,LiuHM:Prb}. Such a metallic ferroelectricity can be characterized according to the structural information, e.g. via measuring convergent-beam electron diffraction or neutron scattering, as done in Ref.~\onlinecite{Shi:Nm}. Physically, a hint for the metallic ferroelectrics is the weak coupling between the electrons at the Fermi level and the (soft) phonon(s) responsible for removing inversion symmetry \cite{Puggioni:Nc,Xiang:Prb,LiuHM:Prb}.

According to the above analysis, the ferroelectric distortion from Y-O-icosahedra should be robust in our $n=2$ superlattice, which in principle can also be expected in other ($AB$O$_3$)$_2$/($AB$'O$_3$)$_2$ multilayers. Following this argument, a promising way to pursuit metallic ferroelectricity is to find a metallic ($AB$O$_3$)$_2$/($AB$'O$_3$)$_2$ superlattices. Still taking (YFeO$_{3}$)$_2$/(YTiO$_{3}$)$_2$ as a model system, most electron bands near the Fermi level are contributed by Fe$^{2+}$ and Ti$^{4+}$, but the structural distortions of Y-O-icosahedra are the first driving force for polarization. Thus, it is expectable that the ferroelectric distortion can survive even \textit{if} the system could become metallic, namely rending a metallic ferroelectricity.

\begin{figure}
\centering
\includegraphics[width=0.45\textwidth]{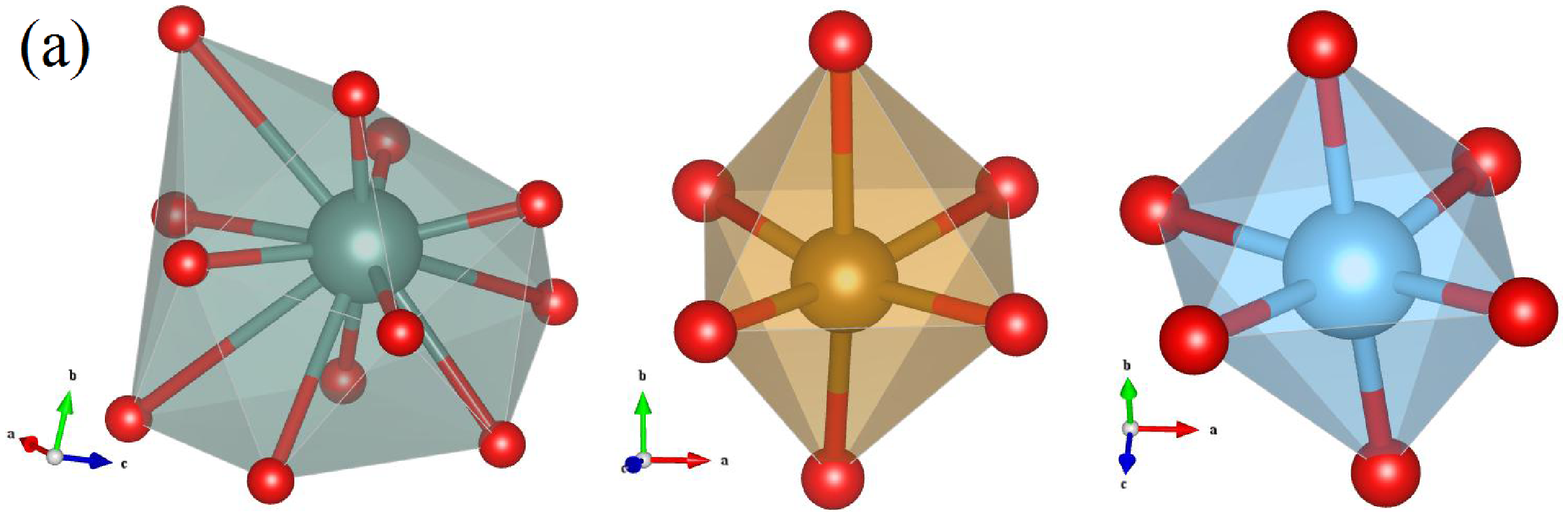}
\includegraphics[width=0.48\textwidth]{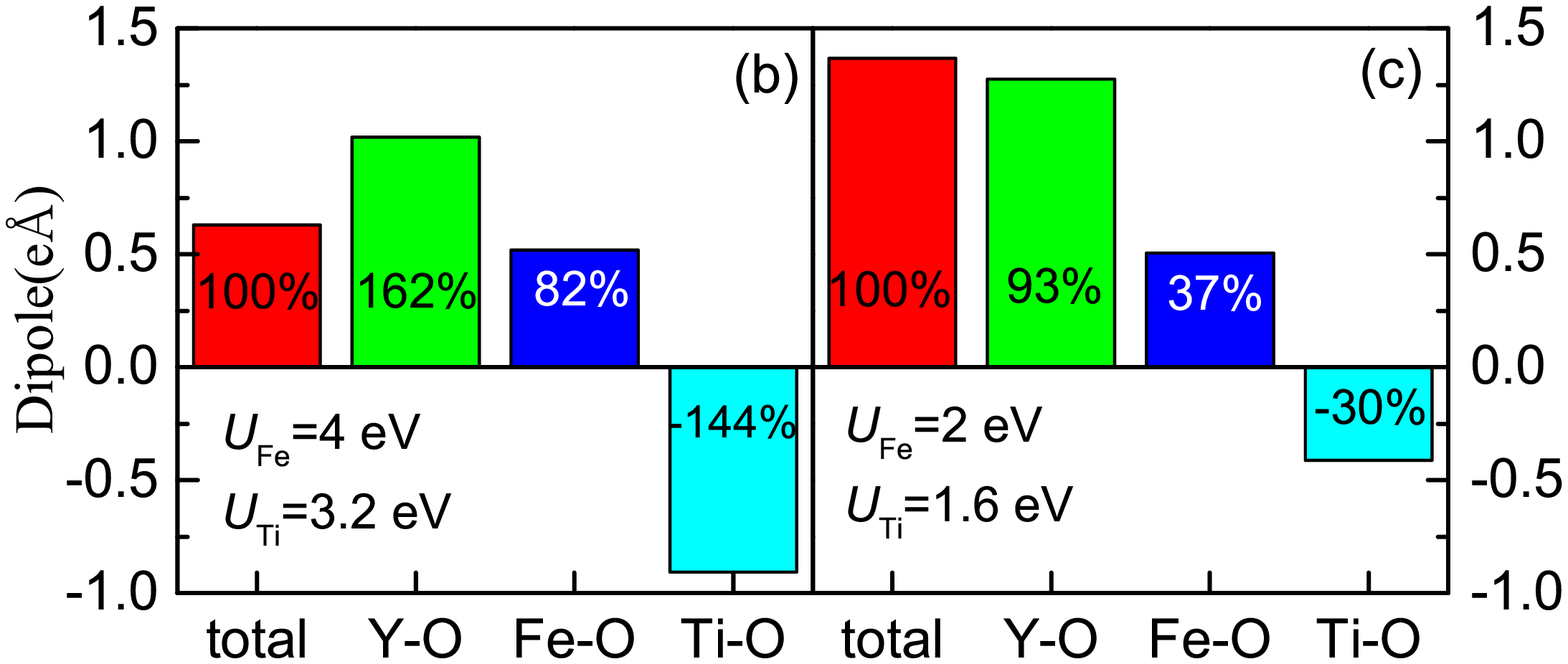}
\caption{(Color online) (a) Sketch of Y-O-icosahedron (left), Fe-O-octahedra (middle), Ti-O-octahedra (right) in the superlattices. (b-c) Point-charge model estimation of individual contribution to total polarization at different $U_{\rm eff}$'s. Fe$^{2+}$-Ti$^{4+}$ is adopted. (b) Insulating. (c) Metallic.}
\label{fe}
\end{figure}

Despite many experimental approaches (e.g. doping, vacancies, or using other perovskites) to make metallic ($AB$O$_3$)$_2$/($AB$'O$_3$)$_2$, here we use smaller $U_{\rm eff}$'s in the DFT calculations to obtain the metallicity. Although smaller $U_{\rm eff}$'s may be not realistic for real YFeO$_3$/YTiO$_3$, the propose of this hypothesis is to illustrate the general physical mechanism that the improper ferroelectricity in ($AB$O$_3$)$_2$/($AB$'O$_3$)$_2$ is robust against the metallicity, going beyond a special property limited to a concrete material (e.g. stoichiometric YFeO$_3$/YTiO$_3$). In the DFT calculation, weak $U_{\rm eff}$'s give rise to finite density of states at the Fermi level although the charge transfer remains (almost) complete. Although the Berry phase method can not work anymore to evaluate the polarization in the metallic state, the point charge model still works.

With decreasing ($U_{\rm Ti}$, $U_{\rm Fe}$), the system turns to be metallic but its ferroelectric distortion can be even more prominent. This enhanced ferroelectricity in metallic state is also understandable: the negative contribution from Ti-O-octahedra is significantly suppressed while the one from Y-O-icosahedra is more robust. Taking the ($U_{\rm Ti}=1.6$ eV, $U_{\rm Fe}=2$ eV) case for example, the contribution from Ti-O-octahedra is only $45\%$ of the original value, while the Fe-O-octahedra contribution is almost unchanged ($97\%$ of the original one), as shown in Fig.~\ref{fe}(c). The contribution of Y-O-icosahedra is slightly increased by $25\%$, which is the largest contribution to the total net polarization. This semi-quantitative partition can help to understand the metallic ferroelectricity.

Our DFT study supports the argument that the improper ferroelectricity due to the geometry factor is robust against the finite density of states at the Fermi level, when the origins of ferroelectric displacement and metallicity are different. Although the $U_{\rm eff}$'s used here may be a little lower for concrete YFeO$_3$/YTiO$_3$, the general physics raised in the present work is scientific sound. Even if the stoichiometric $n=2$ YFeO$_3$/YTiO$_3$ superlattice is not metallic, other approaches can be employed to make the superlattice metallic or other systems can be designed to realize the metallic ferroelectricity following our above argument.

\section{Conclusion}
In summary, the (YFeO$_{3}$)$_{n}$/(YTiO$_{3}$)$_{n}$ superlattices have been studied using the standard DFT calculation. Since the two parent materials are both Mott insulators, unexpected charge transfer has been found, in opposite to the intuitional band alignment scenario. In addition, the ferroelectricity is predicted in the $n=2$ superlattice from the symmetry analysis and confirmed by calculations. Considering the magnetism of Fe, this $n=2$ superlattice is multiferroic. In addition, this ferroelectricity is robust against the metallicity. Even if the real (YFeO$_{3}$)$_2$/(YTiO$_{3}$)$_2$ may be insulating, this design principle can be extended to other superlattices to search for metallic ferroelectrics.

\acknowledgments{Work was supported by the 973 Projects of China (Grant No. 2011CB922101), National Natural Science Foundation of China (Grant Nos. 11274060 and 51322206), the Natural Science Foundation of Jiangsu Province of China (Grant No. BK20141329).}

\bibliographystyle{apsrev4-1}
\bibliography{../../ref}
\end{document}